\documentclass[aps,nofootinbib,preprint,superscriptaddress]{revtex4}%
\usepackage{hyperref}
\usepackage{amsmath}
\usepackage{amsfonts}
\usepackage{amssymb}
\usepackage{graphicx}
\usepackage{color}%
\providecommand{\U}[1]{\protect\rule{.1in}{.1in}}

\begin{document}
\title{The $SL(K+3,%
\mathbb{C}
)$ symmetry of string scatterings from D-branes}
\author{Sheng-Hong Lai}
\email{xgcj944137@gmail.com}
\affiliation{Department of Electrophysics, National Yang Ming Chiao Tung University,
Hsinchu, Taiwan, R.O.C.}
\author{Jen-Chi Lee}
\email{jcclee@cc.nctu.edu.tw}
\affiliation{Department of Electrophysics, National Yang Ming Chiao Tung University,
Hsinchu, Taiwan, R.O.C.}
\author{Yi Yang}
\email{yiyang@mail.nctu.edu.tw}
\affiliation{Department of Electrophysics, National Yang Ming Chiao Tung University,
Hsinchu, Taiwan, R.O.C.}
\date{\today}

\begin{abstract}
By using the solvability of Lauricella function $F_{D}^{(K)}\left(
\alpha;\beta_{1},...,\beta_{K};\gamma;x_{1},...,x_{K}\right)  $ with
nonpositive integer $\beta_{J}$, we show that each scattering or decay process
of string and D-brane states at \textit{arbitrary} mass levels can be
expressed in terms of a single Lauricella function. This result extends the
previous exact $SL(K+3,%
\mathbb{C}
)$ symmetry of tree-level open bosonic string theory to include the D-brane.

\end{abstract}
\maketitle

\section{Introduction}

Motivated by the previous calculation of high energy symmetry
\cite{Gross,Gross1} of string scattering amplitudes (SSA)
\cite{GM,GM1,GrossManes},\cite{ChanLee,ChanLee2,CHL,CHLTY2,CHLTY1}, it was
shown \cite{LSSA} recently that all SSA of four arbitrary string states of the
open bosonic string theory at all kinematic regimes can be expressed in terms
of the $D$-type Lauricella functions $F_{D}^{(K)}\left(  \alpha;\beta
_{1},...,\beta_{K};\gamma;x_{1},...,x_{K}\right)  $ with associated exact
$SL(K+3,%
\mathbb{C}
)$ symmetry (see the definition of $K$ in Eq.(\ref{KKK}). On the other hand, a
class of polarized fermion SSA (PFSSA) at arbitrary mass levels of the
R-sector of the fermionic string theory can also be expressed in terms of the
$D$-type Lauricella functions \cite{pfssa}. Indeed, it can be shown
\cite{LSSA} that these Lauricella functions form an infinite dimensional
representation of the $SL(K+3,%
\mathbb{C}
)$ symmetry group. Moreover, it was demonstrated that there existed $K+2$
recursion relations among the $D$-type Lauricella functions. These recursion
relations can be used to reproduce the Cartan subalgebra and simple root
system of the $SL(K+3,%
\mathbb{C}
)$ group with rank $K+2$ and vice versa.

For the cases of nonpositive integer $\beta_{J}$ (see Eq.(\ref{RR})) in the
$D$-type Lauricella functions which correspond to the cases of the SSA or the
Lauricella SSA (LSSA) mentioned above, the $SL(K+3,%
\mathbb{C}
)$ group or the corresponding $K+2$ stringy Ward identities among the LSSA can
be used to solve \cite{solve} all the LSSA and express them in terms of one
amplitude. These exact Ward identities among the exact LSSA are
generalizations of the \textit{linear relations} with \textit{constant
coefficients} among \ SSA in the hard scattering limit conjectured by Gross
\cite{Gross,Gross1} in 1988 and later corrected and proved in
\cite{ChanLee,ChanLee2,CHL,CHLTY2,CHLTY1}.

More recently, by using the string theory extension \cite{stringbcfw},
\cite{bcfw3,bcfw4} of the field theory BCFW on-shell recursion relations
\cite{bcfw1,bcfw2} , one can show that \cite{LLY3} the residues of all
$n$-point SSA including the Koba-Nielsen (KN) amplitudes can be expressed in
terms of the Lauricella functions with nonpositive integer $\beta_{J}$. As a
result, the above $SL(K+3,%
\mathbb{C}
)$ symmetry group of the $4$-point LSSA was extended to the $n$-point LSSA
with arbitrary $n$ \cite{exact}. It is thus believed that the $SL(K+3,%
\mathbb{C}
)$ symmetry is the fundamental symmetry of the whole bosonic string theory, at
least, tree-level of the bosonic string theory.

To justify the conjecture of the fundamental $SL(K+3,%
\mathbb{C}
)$ symmetry of the bosonic string theory, one needs to collect more evidences
or more SSA to support the proposed exact symmetry. In this paper, we will
show that the scattering and decay processes of string and D-brane states at
arbitrary mass levels can again be expressed in terms of the $D$-type
Lauricella functions with nonpositive integer $\beta_{J}$. This result extends
the previous exact $SL(K+3,%
\mathbb{C}
)$ symmetry of tree-level open bosonic string theory to include the D-brane.
This is also consistent with the previous results that the linear relations
with constant coefficients among\ SSA in the hard scattering limit persist for
the processes of D-brane scatterings \cite{DD} and decays \cite{decay} as they
are all related to the exact $SL(K+3,%
\mathbb{C}
)$ symmetry of bosonic string theory. We will see that the calculation will be
greatly simplified by using the solvability \cite{solve} of the Lauricella
functions with nonpositive integer $\beta_{J}$.

\section{Review of $SL(K+3,C)$ symmetry}

We first review the LSSA of three tachyons and one arbitrary string states in
the $26D$ open bosonic string theory and its associated $SL(K+3,%
\mathbb{C}
)$ symmetry. The general states are of the following form \cite{LSSA}%
\begin{equation}
\left\vert r_{n}^{T},r_{m}^{P},r_{l}^{L}\right\rangle =\prod_{n>0}\left(
\alpha_{-n}^{T}\right)  ^{r_{n}^{T}}\prod_{m>0}\left(  \alpha_{-m}^{P}\right)
^{r_{m}^{P}}\prod_{l>0}\left(  \alpha_{-l}^{L}\right)  ^{r_{l}^{L}}%
|0,k\rangle\label{state}%
\end{equation}
where $e^{P}=\frac{1}{M_{2}}(E_{2},\mathrm{k}_{2},0)=\frac{k_{2}}{M_{2}}$ is
the momentum polarization, $e^{L}=\frac{1}{M_{2}}(\mathrm{k}_{2},E_{2},0)$ is
the longitudinal polarization and $e^{T}=(0,0,1)$ is the transverse
polarization on the $\left(  2+1\right)  $-dimensional scattering plane. In
addition to the mass level $M_{2}^{2}=2(N-1)$ with%
\begin{equation}
N=\sum_{\substack{n,m,l>0\\\{\text{ }r_{j}^{X}\neq0\}}}\left(  nr_{n}%
^{T}+mr_{m}^{P}+lr_{l}^{L}\right)  ,\label{NN}%
\end{equation}
we define another important index $K$ for the state in Eq.(\ref{state})%
\begin{equation}
K=\sum_{\substack{n,m,l>0\\\{\text{ }r_{j}^{X}\neq0\}}}\left(  n+m+l\right)
\label{KKK}%
\end{equation}
where $X=\left(  T,P,L\right)  $ and we have put $r_{n}^{T}=r_{m}^{P}%
=r_{l}^{L}=1$ in Eq.(\ref{NN}) in the definition of $K$. Intuitively, $K$
counts the number of variaty of the $\alpha_{-j}^{X}$ oscillators in
Eq.(\ref{state}). For later use, we also define%
\begin{equation}
k_{j}^{X}\equiv e^{X}\cdot k_{j}\text{ \ for \ }X=\left(  T,P,L\right)  .
\end{equation}
Note that SSA of three tachyons and one arbitrary string states with
polarizations orthogonal to the scattering plane \textit{vanish}.

Note that to achieve BRST invariance or physical state conditions in the old
covariant quantization scheme for the state in Eq.(\ref{state}), one needs to
add polarizations and put on the Virasoro constraints. As an example, let's
calculate the case of symmetric spin $3$ state of mass level $M_{2}^{2}=4$. We
first note that the three momentum polarizations defined on the scattering
plane above satisfy the completeness relation%

\begin{equation}
\eta^{\mu\nu}=\underset{\alpha,\beta}{%
{\textstyle\sum}
}e_{\alpha}^{\mu}e_{\beta}^{\nu}\eta^{\alpha\beta} \label{2.11.}%
\end{equation}
where $\mu,\nu=0,1,2$ and $\alpha,\beta=P,L,T.$ $Diag$ $\eta^{\mu\nu
}=(-1,1,1)$. We can use Eq.(\ref{2.11.}) to transform all $\mu,\nu$
coordinates to coordinates $\alpha,\beta$ on the scattering plane. One gauge
choice of the symmetric spin $3$ state with Virasoro constraints can be
calculated to be%

\begin{equation}
\epsilon_{\mu\nu\lambda}\alpha_{-1}^{\mu\nu\lambda}\left\vert 0,k\right\rangle
;k^{\mu}\epsilon_{\mu\nu\lambda}=0,\eta^{\mu\nu}\epsilon_{\mu\nu\lambda}=0.
\label{2.22..}%
\end{equation}
We can then use the helicity decomposition and writing $\epsilon_{\mu
\nu\lambda}=\Sigma_{\mu,\nu,\lambda}e_{\mu}^{\alpha}e_{\nu}^{\beta}e_{\lambda
}^{\delta}u_{\alpha\beta\delta};\alpha,\beta,\delta=P,L,T$ to get%

\begin{equation}
\epsilon_{\mu\nu\lambda}\alpha_{-1}^{\mu\nu\lambda}\left\vert 0,k\right\rangle
=[u_{TTL}(3\alpha_{-1}^{TTL}-\alpha_{-1}^{LLL})+u_{TTT}(\alpha_{-1}%
^{TTT}-3\alpha_{-1}^{LLT})]\left\vert 0,k\right\rangle . \label{2.23..}%
\end{equation}
It is now easy to see from Eq.(\ref{2.23..}) that to achieve BRST invariance
the spin $3$ state can be written as a linear combination of states in
Eq.(\ref{state}) with coefficients $u_{TTL}$ and $u_{TTT}$.

The 4-point LSSA of three tachyons and one string state in Eq.(\ref{state})
can be calculated to be \cite{LSSA}%
\begin{equation}
A_{4}=B\left(  -\frac{t}{2}-1,-\frac{s}{2}-1\right)  F_{D}^{(K)}\left(
-\frac{t}{2}-1;R_{n}^{X};\frac{u}{2}+2-N;\tilde{Z}_{n}^{X}\right)  \prod
_{X}\left(  \prod_{n=1}\left[  -(n-1)!k_{3}^{X}\right]  ^{r_{n}^{X}}\right)
\label{st}%
\end{equation}
where $B(a,b)$ is the Beta function with $\left(  s,t\right)  $ being the
usual Mandelstam variables, $k_{i}^{X}$ is the momentum of the $i$th string
state projected on the $X$ polarization. In Eq.(\ref{st}), we have defined%
\begin{equation}
R_{l}^{X}\equiv\left\{  -r_{1}^{X}\right\}  ^{1},\cdots,\left\{  -r_{l}%
^{X}\right\}  ^{l}\text{ \ with \ }\left\{  a\right\}  ^{n}%
=\underset{n}{\underbrace{a,a,\cdots,a}}.\label{RR}%
\end{equation}
for the $\beta_{J}$ in the Lauricella function and%
\begin{equation}
Z_{l}^{X}\equiv\left[  z_{1}^{X}\right]  ,\cdots,\left[  z_{l}^{X}\right]
\text{ \ \ with \ \ }\left[  z_{l}^{X}\right]  =z_{l0}^{X},\cdots,z_{l\left(
l-1\right)  }^{X}.\label{comp}%
\end{equation}
where in Eq.(\ref{comp}), we have defined%
\begin{equation}
z_{l}^{X}=\left\vert \left(  -\frac{k_{1}^{X}}{k_{3}^{X}}\right)  ^{\frac
{1}{k}}\right\vert ,\ z_{ll^{\prime}}^{X}=z_{l}^{X}e^{\frac{2\pi il^{\prime}%
}{l}},\ \tilde{z}_{ll^{\prime}}^{X}\equiv1-z_{ll^{\prime}}^{X}\text{ \ \ for
\ \ }l^{\prime}=0,\cdots,l-1.
\end{equation}
It is important to note that all $\beta_{j}$ of $F_{D}^{(K)}$ in Eq.(\ref{st})
are nonpositive integer. The $D$-type Lauricella function $F_{D}^{(K)}$ in
Eq.(\ref{st}) is defined to be%
\begin{equation}
F_{D}^{(K)}\left(  \alpha;\beta_{1},...,\beta_{K};\gamma;x_{1},...,x_{K}%
\right)  =\sum_{n_{1},\cdots,n_{K}=0}^{\infty}\frac{\left(  \alpha\right)
_{n_{1}+\cdots+n_{K}}}{\left(  \gamma\right)  _{n_{1}+\cdots+n_{K}}}%
\frac{\left(  \beta_{1}\right)  _{n_{1}}\cdots\left(  \beta_{K}\right)
_{n_{K}}}{n_{1}!\cdots n_{K}!}x_{1}^{n_{1}}\cdots x_{K}^{n_{K}}%
\end{equation}
where $(\alpha)_{n}=\alpha\cdot\left(  \alpha+1\right)  \cdots\left(
\alpha+n-1\right)  $ is the Pochhammer symbol. The result in Eq.(\ref{st}) can
be generalized to LSSA of four arbitrary string states, and to those of $n$
arbitrary string states (see Eq.(\ref{pro})) below \cite{LLY3,exact}.

For illustration, we calculate the Lauricella functions which correspond to
the LSSA for levels $K=1,2$. For $K=1$, there are three type of LSSA
$(\alpha=-\frac{t}{2}-1,\gamma=\frac{u}{2}+2)$%
\begin{align}
(\alpha_{-1}^{T})^{p_{1}}\text{, }F_{D}^{(1)}(\alpha,-p_{1},\gamma
-p_{1},1)\text{, }N  &  =p_{1},\\
(\alpha_{-1}^{P})^{q_{1}}\text{, }F_{D}^{(1)}(\alpha,-q_{1},\gamma
-q_{1},\left[  \tilde{z}_{1}^{P}\right]  )\text{, }N  &  =q_{1},\\
(\alpha_{-1}^{L})^{r_{1}}\text{, }F_{D}^{(1)}(\alpha,-r_{1},\gamma
-r_{1},\left[  \tilde{z}_{1}^{L}\right]  )\text{, }N  &  =r_{1}.
\end{align}
For $K=2$, there are six type of LSSA%
\begin{align}
(\alpha_{-1}^{T})^{p_{1}}(\alpha_{-1}^{P})^{q_{1}}\text{, }F_{D}^{(2)}%
(\alpha,-p_{1},-q_{1},\gamma-p_{1}-q_{1},1,\left[  \tilde{z}_{1}^{P}\right]
)\text{,}N  &  =p_{1}+q_{1},\\
(\alpha_{-1}^{T})^{p_{1}}(\alpha_{-1}^{L})^{r_{1}}\text{, }F_{D}^{(2)}%
(\alpha,-p_{1},-r_{1},\gamma-p_{1}-r_{1},1,\left[  \tilde{z}_{1}^{L}\right]
)\text{,}N  &  =p_{1}+r_{1},\\
(\alpha_{-1}^{P})^{q_{1}}(\alpha_{-1}^{L})^{r_{1}}\text{, }F_{D}^{(2)}%
(\alpha,-q_{1},-r_{1},\gamma-q_{1}-r_{1},\left[  \tilde{z}_{1}^{P}\right]
,\left[  \tilde{z}_{1}^{L}\right]  )\text{,}N  &  =q_{1}+r_{1},\\
(\alpha_{-2}^{T})^{p_{2}}\text{, }F_{D}^{(2)}(\alpha,-p_{2},-p_{2}%
,\gamma-2p_{2},1,1)\text{, }N  &  =2p_{2},\\
(\alpha_{-2}^{P})^{q_{2}}\text{, }F_{D}^{(2)}(\alpha,-q_{2},-q_{2}%
,\gamma-2q_{2},1-Z_{2}^{P},1-\omega Z_{2}^{P})\text{, }N  &  =2q_{2},\\
(\alpha_{-2}^{L})^{r_{2}}\text{, }F_{D}^{(2)}(\alpha,-r_{2},-r_{2}%
,\gamma-2r_{2},1-Z_{2}^{L},1-\omega Z_{2}^{L})\text{, }N  &  =2r_{2}.
\end{align}
It is important to note that for a given $K$, there are infinite number of
string states with arbitrary higher mass levels. Moreover, each string state
was assigned a particular value of integer $K$, and its associated LSSA is a
basis of the $SL(K+3,%
\mathbb{C}
)$ group representation.

To demonstrate the $SL(K+3,%
\mathbb{C}
)$ symmetry of the LSSA, one first defines the basis functions \cite{slkc}%
\begin{align}
&  f_{ac}^{b_{1}\cdots b_{K}}\left(  \alpha;\beta_{1},\cdots,\beta_{K}%
;\gamma;x_{1},\cdots,x_{K}\right) \nonumber\\
&  =B\left(  \gamma-\alpha,\alpha\right)  F_{D}^{\left(  K\right)  }\left(
\alpha;\beta_{1},\cdots,\beta_{K};\gamma;x_{1},\cdots,x_{K}\right)  a^{\alpha
}b_{1}^{\beta_{1}}\cdots b_{K}^{\beta_{K}}c^{\gamma}, \label{id2}%
\end{align}
so that the LSSA in Eq.(\ref{st}) can be rewritten as \cite{Group}
\begin{equation}
A_{4}=f_{11}^{-(n-1)!k_{3}^{X}}\left(  -\frac{t}{2}-1;R_{n}^{X},;\frac{u}%
{2}+2-N;\tilde{Z}_{n}^{X}\right)  .
\end{equation}
One can then introduce the $(K+3)^{2}-1$ generators $\mathcal{E}_{ij}$ of
$SL(K+3,%
\mathbb{C}
)$ group \cite{slkc,Group}%
\begin{equation}
\left[  \mathcal{E}_{ij},\mathcal{E}_{kl}\right]  =\delta_{jk}\mathcal{E}%
_{il}-\delta_{li}\mathcal{E}_{kj};\text{ \ }1\leqslant i,j\leqslant K+3
\end{equation}
to operate on the basis functions in Eq.(\ref{id2}). These are $1$ $E^{\alpha
}$, $K$ $E^{\beta_{k}}(k=1,2\cdots K)$, $1$ $E^{\gamma}$,$1$ $E^{\alpha\gamma
}$, $K$ $E^{\beta_{k}\gamma}$ and $K$ $E^{\alpha\beta_{k}\gamma}$ which sum up
to $3K+3$ raising generators. There are also $3K+3$ corresponding lowering
operators. In addition, there are $K\left(  K-1\right)  $ $E_{\beta_{p}%
}^{\beta_{k}}$ and $K+2$ $\left\{  J_{\alpha},J_{\beta_{k}},J_{\gamma
}\right\}  $ generators, the Cartan subalgebra. In sum, the total number of
generators are $2(3K+3)+K(K-1)+K+2=(K+3)^{2}-1$ \cite{Group}.

For the general $4$-point LSSA, it is straightforward to calculate the
operation of these generators on the basis functions and show the $SL(K+3,%
\mathbb{C}
)$ symmetry \cite{Group}. For the cases of higher point $(n\geq5)$ LSSA, one
encounters the operation on the sum of products of the Lauricella functions
\cite{LLY3}%
\begin{equation}
\text{Residue of }n\text{-point }LSSA\sim\sum\text{coefficient}\prod
(\text{single tensor }4\text{-point }LSSA). \label{pro}%
\end{equation}
Therefore, one needs to deal with product representations of $SL(K+3,%
\mathbb{C}
)$.

Indeed, we have recently applied the string theory extension of field theory
BCFW on-shell recursion relations \cite{bcfw1,bcfw2} to show that the
$SL(K+3,C)$ symmetry group of the $4$-point LSSA persists for general
$n$-point SSA with arbitrary higher point couplings in string theory
\cite{exact}. We thus have shown that the $SL(K+3,C)$ symmetry is an exact
symmetry of the whole bosonic string theory, and that all $n$-point SSA of the
bosonic string theory form an infinite dimensional representation of the
$SL(K+3,C)$ group. Moreover, all residues of SSA in the string theory on-shell
recursion prescription can be expressed in terms of the four-point LSSA.

There is an interesting issue of the stringy on-shell Ward identities or
decoupling of zero-norm states associated with the $SL(K+3,C)$ symmetry. For
the $n$-point Ward identities with $n\geq5$, one can either write down the
linear on-shell Ward identities in terms of $n$-point functions or, through
reduction of stringy BCFW recursion, calculate the non-linear on-shell Ward
identities in terms of $4$-point functions. For the latter case, we conjecture
that the non-linear Ward identities can be reduced to the equivalent linear
$4$-point Ward identities since both forms of Ward identities are associated
with the same $SL(K+3,C)$ group.

\section{Solvability of LSSA}

There exist $K+2$ recurrence relations for the $D$-type Lauricella functions
\cite{Group}. Moreover, these recurrence relations can be used to reproduce
the Cartan subalgebra and simple root system of the $SL(K+3,%
\mathbb{C}
)$ group with rank $K+2$ \cite{Group}. With the Cartan subalgebra and the
simple roots, one can easily write down the whole Lie algebra of the $SL(K+3,%
\mathbb{C}
)$ group. So one can construct the $SL(K+3,%
\mathbb{C}
)$ Lie algebra from the recurrence relations and vice versa.

On the other hand, one can use the $K+2$ recurrence relations to deduce the
following key recurrence relation \cite{solve}%
\begin{align}
&  x_{j}F_{D}^{(K)}\left(  \alpha;\beta_{1},.,\beta_{i}-1,..,\beta_{K}%
;\gamma;x_{1},...,x_{K}\right)  -x_{i}F_{D}^{(K)}\left(  \alpha;\beta
_{1},.,\beta_{j}-1,..,\beta_{K};\gamma;x_{1},...,x_{K}\right) \nonumber\\
&  +\left(  x_{i}-x_{j}\right)  F_{D}^{(K)}\left(  \alpha;\beta_{1}%
,...,\beta_{K};\gamma;x_{1},...,x_{K}\right)  =0, \label{key}%
\end{align}
which, for the case of nonpositive $\beta_{j}$, can be repeatedly used to
decrease the value of $K$ and reduce all the Lauricella functions $F_{D}%
^{(K)}$ in the LSSA to the Gauss hypergeometric functions $F_{D}^{(1)}=$
$_{2}F_{1}(\alpha,\beta,\gamma,x)$. Indeed, One can repeatedly apply
Eq.(\ref{key}) to the Lauricella functions in Eq.(\ref{st}) and express
$F_{D}^{(K)}\left(  \alpha;\beta_{1},...,\beta_{K};\gamma;x_{1},...,x_{K}%
\right)  $ in terms of $F_{D}^{(K-1)}(\beta_{1},..\beta_{i-1},\beta
_{i+1}...\beta_{j}^{\prime},...\beta_{K})$ with $\beta_{j}^{\prime}=\beta
_{j},\beta_{j}-1,...,\beta_{j}-\left\vert \beta_{i}\right\vert $ or
$F_{D}^{(K-1)}(\beta_{1},...\beta_{i}^{\prime},...\beta_{j-1},\beta
_{j+1},...\beta_{K})$ with $\beta_{i}^{\prime}=\beta_{i},\beta_{i}%
-1,...,\beta_{i}-\left\vert \beta_{j}\right\vert $ (assuming $i<j$).

For example, for say $K=2$, Eq.(\ref{key}) reduces to%
\begin{align}
&  x_{2}F_{D}^{(2)}\left(  \alpha;\beta_{1}-1,\beta_{2};\gamma;x_{1}%
,x_{2}\right)  -x_{1}F_{D}^{(2)}\left(  \alpha;\beta_{1},\beta_{2}%
-1;\gamma;x_{1},x_{2}\right) \nonumber\\
&  +\left(  x_{1}-x_{2}\right)  F_{D}^{(2)}\left(  \alpha;\beta_{1},\beta
_{2};\gamma;x_{1},x_{2}\right)  =0.
\end{align}

For say $\beta_{1}=0$ and $\beta_{2}=-1$, we get%
\begin{align}
x_{2}F_{D}^{(2)}\left(  \alpha;-1,-1;\gamma;x_{1},x_{2}\right)   &
=x_{1}F_{D}^{(2)}\left(  \alpha;0,-2;\gamma;x_{1},x_{2}\right)  -\left(
x_{1}-x_{2}\right)  F_{D}^{(2)}\left(  \alpha;0,-1;\gamma;x_{1},x_{2}\right)
\nonumber\\
&  =x_{1}F_{D}^{(1)}\left(  \alpha;-2;\gamma;x_{2}\right)  -\left(
x_{1}-x_{2}\right)  F_{D}^{(1)}\left(  \alpha;-1;\gamma;x_{2}\right)
\end{align}
which express the Lauricella function with $K=2$ in terms of those of $K=1$.
We can repeat similar process to decrease the value of $K$.

Moreover, one can further reduce the Gauss hypergeometric functions by
deriving a multiplication theorem for them, and then solve \cite{solve} all
the LSSA in terms of one single amplitude. This solvability is crucial to show
that all scattering and decay processes of string and D-brane states at
arbitrary mass levels can be expressed in terms of the Lauricella function and
thus its associated $SL(K+3,%
\mathbb{C}
)$ symmetry.

\section{Closed string scattered off D-brane}

In this paper, we will consider scattering and decay processes of string and
D-brane states at arbitrary mass levels. These are three classes of processes
\cite{Klebanov}, \cite{D2,D3,D4,D5,D6}: (A). Closed string scattered off
D-brane, (B). Closed string decays into two open strings on the brane and (C).
Four open string scattering on the brane. The calculation of process (C) is
similar to that of four open string scattering without D-brane, and thus can
be expressed in terms of the Lauricella function.

In this section, we first consider process (A). In \cite{Klebanov}, the
calculation was done only for the massless string states. Here we will
consider scatterings of arbitrary massive string states for the bosonic
string. The standard propagators of the left and right moving fields are
$\left\langle X^{\mu}\left(  z\right)  X^{\nu}\left(  w\right)  \right\rangle
=-\eta^{\mu\nu}\log\left(  z-w\right)  ,\left\langle \tilde{X}^{\mu}\left(
\bar{z}\right)  \tilde{X}^{\nu}\left(  \bar{w}\right)  \right\rangle
=-\eta^{\mu\nu}\log\left(  \bar{z}-\bar{w}\right)  .$ In addition, there are
nontrivial correlator as well \cite{Klebanov}%
\begin{equation}
\left\langle X^{\mu}\left(  z\right)  \tilde{X}^{\nu}\left(  \bar{w}\right)
\right\rangle =-D^{\mu\nu}\log\left(  z-\bar{w}\right)  \label{DD}%
\end{equation}
as a result of the Dirichlet boundary condition at the real axis. The diagonal
matrix $D$ in Eq.(\ref{DD}) reverses the sign for fields satisfying Dirichlet
boundary condition. That is, there are $p+1$ Neumann and $25-p$ Dirichlet for
a Dp-brane. We will follow the standard notation and make the following
replacement \cite{Klebanov}%
\begin{equation}
\tilde{X}^{\mu}\left(  \bar{z}\right)  \rightarrow D_{\text{ \ }\nu}^{\mu
}X^{\nu}\left(  \bar{z}\right)
\end{equation}
which allows us to use the standard correlators throughout our calculations.
As a warm up exercise, we first consider tachyon to tachyon scattering
\cite{DD}%
\begin{align}
A_{tach}  &  =\int d^{2}z_{1}d^{2}z_{2}\left\langle V_{1}\left(  z_{1},\bar
{z}_{1}\right)  V_{2}\left(  z_{2},\bar{z}_{2}\right)  \right\rangle
\nonumber\\
&  =\int d^{2}z_{1}d^{2}z_{2}\left(  z_{1}-\bar{z}_{1}\right)  ^{k_{1}\cdot
D\cdot k_{1}}\left(  z_{2}-\bar{z}_{2}\right)  ^{k_{2}\cdot D\cdot k_{2}%
}\left\vert z_{1}-z_{2}\right\vert ^{2k_{1}\cdot k_{2}}\left\vert z_{1}%
-\bar{z}_{2}\right\vert ^{2k_{1}\cdot D\cdot k_{2}}.
\end{align}
To fix the $SL\left(  2,R\right)  $ invariance, we set $z_{1}=iy$ and
$z_{2}=i$ and, for the contribution of the $(0,1)$ interval, we obtain
\cite{DD}
\begin{align}
A_{tach}^{(0,1)}  &  =4\left(  2i\right)  ^{2a_{0}}\int_{0}^{1}dy\text{
}y^{a_{0}}\left(  1-y\right)  ^{b_{0}}\left(  1+y\right)  ^{c_{0}}\nonumber\\
&  =4\left(  2i\right)  ^{2a_{0}}2^{-2a_{0}-1-N}\int_{0}^{1}dx\text{ }%
x^{b_{0}}\left(  1-x\right)  ^{a_{0}}\left(  1+x\right)  ^{a_{0}+N}.
\label{ta5}%
\end{align}
In the above calculations, we have defined
\begin{align}
a_{0}  &  =k_{1}\cdot D\cdot k_{1}=k_{2}\cdot D\cdot k_{2},\\
b_{0}  &  =2k_{1}\cdot k_{2}+1,\\
c_{0}  &  =2k_{1}\cdot D\cdot k_{2}+1,
\end{align}
so that%
\begin{equation}
2a_{0}+b_{0}+c_{0}+2=4N_{1}\equiv-N,
\end{equation}
and $-k_{1}^{2}=M^{2}\equiv\frac{\alpha_{closed}^{\prime}M_{closed}^{2}}%
{2}=2(N_{1}-1)$, $N_{1}=0$ for tachyon. The momentum conservation on the
D-brane%
\begin{equation}
D\cdot k_{1}+k_{1}+D\cdot k_{2}+k_{2}=0 \label{con}%
\end{equation}
is crucial to get the final result Eq.(\ref{ta5}). Similarly, For the
contribution of the $(1,\infty)$ interval, we end up with%
\begin{align}
A_{tach}^{(1,\infty)}  &  =4\left(  2i\right)  ^{2a_{0}}\int_{1}^{\infty
}dy\text{ }y^{a_{0}}\left(  y-1\right)  ^{b_{0}}\left(  1+y\right)  ^{c_{0}%
}\nonumber\\
&  =4\left(  2i\right)  ^{2a_{0}}2^{-2a_{0}-1-N}\int_{0}^{1}dx\text{ }%
x^{b_{0}}\left(  1-x\right)  ^{a_{0}+N}\left(  1+x\right)  ^{a_{0}}.
\end{align}

For the general massive tensor to another massive tensor scattering, the
calculation will be very complicated as there are many new contraction terms.
We will use the \textit{solvability} of the LSSA discussed above to simplify
the calculation. The strategy is as follows: We can simply calculate a typical
term of a given process. If the result turns out to be a Lauricella function
with nonpositive $\beta_{j}$, we can then use the solvability property to
argue that the final amplitude after summing up all typical terms of the
process is a LSSA. To do the calculation, we first define%
\begin{align}
a  &  =k_{1}\cdot D\cdot k_{1}+n_{a},\\
b  &  =2k_{1}\cdot k_{2}+1+n_{b},\\
c  &  =2k_{1}\cdot D\cdot k_{2}+1+n_{c},
\end{align}
where $n_{a}$, $n_{b}$ and $n_{c}$ are integer and then define $N^{\prime
}=-\left(  2n_{a}+n_{b}+n_{c}\right)  $, so that%
\begin{equation}
2a+b+c+2+N^{\prime}=4N_{1}\Longrightarrow2a+b+c+2=4N_{1}-N^{\prime}\equiv-N
\label{int}%
\end{equation}
where $k_{1}^{2}=2(N_{1}-1)$ and $N_{1}$ is the mass level of $k_{1}$. It is
easy to see that a typical term in the general tensor to tensor scattering can
be calculated to be \cite{DD}%

\begin{align}
I_{\left(  0,1\right)  }  &  =\int_{0}^{1}dt\text{ }y^{a}\left(  1-y\right)
^{b}\left(  1+y\right)  ^{c},\nonumber\\
&  =2^{-2a-1-N}\int_{0}^{1}dx\text{ }x^{b}\left(  1-x\right)  ^{a}\left(
1+x\right)  ^{a+N}.
\end{align}
Similarly, for the $(1,\infty)$ interval, one gets
\begin{align}
I_{\left(  1,\infty\right)  }  &  =\int_{1}^{\infty}dy\text{ }y^{a}\left(
y-1\right)  ^{b}\left(  1+y\right)  ^{c}\nonumber\\
&  =2^{-2a-1-N}\int_{0}^{1}dx\text{ }x^{b}\left(  1-x\right)  ^{a+N}\left(
1+x\right)  ^{a}.
\end{align}
The sum of the two channels gives \cite{DD}%
\begin{align}
I  &  =2^{-2a-2-N}\sum_{m=0}^{N}\left[  1+\left(  -1\right)  ^{m}\right]
\binom{N}{m}\cdot\frac{\Gamma\left(  a+1\right)  \Gamma\left(  \frac{b+1}%
{2}+\frac{m}{2}\right)  }{\Gamma\left(  a+\frac{b+3}{2}+\frac{m}{2}\right)
}\nonumber\\
&  =2^{-2a-1-N}\frac{\Gamma\left(  a+1\right)  \Gamma\left(  \frac{b+1}%
{2}\right)  }{\Gamma\left(  a+\frac{b+3}{2}\right)  }\sum_{n=0}^{\left[
\frac{N}{2}\right]  }\binom{N}{2n}\dfrac{\left(  \frac{b+1}{2}\right)  _{n}%
}{\left(  a+\frac{b+3}{2}\right)  _{n}}\nonumber\\
&  =2^{-2a-1-N}\cdot B\left(  a+1,\frac{b+1}{2}\right)  \cdot\text{ }_{3}%
F_{2}\left(  \frac{b+1}{2},-\left[  \frac{N}{2}\right]  ,\dfrac{1}{2}-\left[
\frac{N}{2}\right]  ;a+\frac{b+3}{2},\dfrac{1}{2};1\right)  \label{master}%
\end{align}
where $_{3}F_{2}$ is a generalized hypergeometric function. For the special
arguments of $_{3}F_{2}$ in Eq.(\ref{master}), the hypergeometric function
terminates to a finite sum and, as a result, the whole scattering amplitudes
consistently reduce to the usual beta function. In calculating
Eq.(\ref{master}), we have used the identity%
\begin{equation}
\overset{\left[  \frac{N}{2}\right]  }{\underset{n=0}{\sum}}\binom{N}{2n}%
\frac{\left(  A\right)  _{n}}{\left(  C\right)  _{n}}=\text{ }_{3}F_{2}\left(
A,-\left[  \frac{N}{2}\right]  ,\frac{1}{2}-\left[  \frac{N}{2}\right]
;C,\frac{1}{2};1\right)  , \label{ac}%
\end{equation}
which can be easily proved.

At this point, one might think that the amplitude calculated in
Eq.(\ref{master}) is not a LSSA, and the $SL(K+3,%
\mathbb{C}
)$ group may be just a subgroup of an unknown larger symmetry group
$G\supseteq$ $SL(K+3,%
\mathbb{C}
)$ of the bosonic string theory. However, we will see that this is not the
case. To show that $_{3}F_{2}$ in the amplitude Eq.(\ref{master}) is a LSSA,
we first do a change of variable $y=x^{2}$ to get
\begin{align}
I  &  =2^{-2a-1-N}\int_{0}^{1}dx\text{ }x^{b}\left(  1-x\right)  ^{a}\left(
1+x\right)  ^{a}\left[  \left(  1+x\right)  ^{N}+\left(  1-x\right)
^{N}\right] \nonumber\\
&  =2^{-2a-1-N}\int_{0}^{1}dy\text{ }y^{\frac{b-1}{2}}\left(  1-y\right)
^{a}G\left(  y\right)
\end{align}
where%
\begin{equation}
G\left(  y\right)  =G\left(  x^{2}\right)  =\frac{1}{2}\left[  \left(
1+x\right)  ^{N}+\left(  1-x\right)  ^{N}\right]  .
\end{equation}
We can solve
\begin{equation}
G\left(  x^{2}\right)  =\frac{1}{2}\left[  \left(  1+x\right)  ^{N}+\left(
1-x\right)  ^{N}\right]  =0
\end{equation}
to get%
\begin{equation}
x=x_{k}=\frac{\left(  -1\right)  ^{\frac{1}{N}}-1}{\left(  -1\right)
^{\frac{1}{N}}+1}=\frac{w_{N,k}-1}{w_{N,k}+1}=i\tan\frac{\theta_{k}}{2}
\label{root}%
\end{equation}
where
\begin{equation}
w_{N,k}=e^{\frac{i\pi}{N}+\frac{2i\pi k}{N}}=e^{i\theta_{k}},\theta_{k}%
=\frac{\pi}{N}+\frac{2\pi k}{N},k=1,2\cdots,N.
\end{equation}
We can now do the following factorization
\begin{align}
G\left(  x^{2}\right)   &  =\underset{k=1}{\overset{N}{%
{\displaystyle\prod}
}}\left(  1-\frac{x}{x_{k}}\right)  =\underset{k=1}{\overset{\left[  \frac
{N}{2}\right]  }{%
{\displaystyle\prod}
}}\left(  1-\frac{x}{x_{k}}\right)  \left(  1-\frac{x}{\bar{x}_{k}}\right)
=\underset{k=1}{\overset{\left[  \frac{N}{2}\right]  }{%
{\displaystyle\prod}
}}\left(  1-\left(  \frac{x}{x_{k}}+\frac{x}{\bar{x}_{k}}\right)  +\frac
{x^{2}}{x_{k}\bar{x}_{k}}\right) \nonumber\\
&  =\underset{k=1}{\overset{\left[  \frac{N}{2}\right]  }{%
{\displaystyle\prod}
}}\left(  1+\frac{x^{2}}{x_{k}\bar{x}_{k}}\right)
=\underset{k=1}{\overset{\left[  \frac{N}{2}\right]  }{%
{\displaystyle\prod}
}}\left(  1-\frac{x^{2}}{x_{k}^{2}}\right)
\end{align}
to obtain%
\begin{equation}
G\left(  y\right)  =\underset{k=1}{\overset{\left[  \frac{N}{2}\right]  }{%
{\displaystyle\prod}
}}\left(  1-\frac{y}{x_{k}^{2}}\right)  =\underset{k=1}{\overset{\left[
\frac{N}{2}\right]  }{%
{\displaystyle\prod}
}}\left(  1+\frac{y}{\left(  \tan\frac{\theta_{k}}{2}\right)  ^{2}}\right)
=\underset{k=1}{\overset{\left[  \frac{N}{2}\right]  }{%
{\displaystyle\prod}
}}\left(  1+\frac{y}{\left(  \tan\left(  \frac{\pi}{2N}+\frac{\pi k}%
{N}\right)  \right)  ^{2}}\right)  .
\end{equation}
Finally, we can use%
\begin{align}
&  \int_{0}^{1}dy\text{ }y^{\frac{b-1}{2}}\left(  1-y\right)  ^{a}G\left(
y\right) \nonumber\\
&  =\int_{0}^{1}dy\text{ }y^{\frac{b-1}{2}}\left(  1-y\right)  ^{a}%
\underset{k=1}{\overset{\left[  \frac{N}{2}\right]  }{%
{\displaystyle\prod}
}}\left(  1-\frac{y}{x_{k}^{2}}\right) \nonumber\\
&  =B\left(  a+1,\frac{b+1}{2}\right)  \cdot F_{D}^{\left(  \left[  \frac
{N}{2}\right]  \right)  }\left(  \frac{b+1}{2};\underset{\left[  \frac{N}%
{2}\right]  }{\underbrace{-1,...,-1}};a+\frac{b+3}{2};\frac{1}{x_{1}^{2}%
},...,\frac{1}{x_{\left[  \frac{N}{2}\right]  }^{2}}\right)
\end{align}
to obtain the identification%
\begin{equation}
\text{ }_{3}F_{2}\left(  \frac{b+1}{2},-\left[  \frac{N}{2}\right]  ,\frac
{1}{2}-\left[  \frac{N}{2}\right]  ;a+\frac{b+3}{2},\frac{1}{2};1\right)
=F_{D}^{\left(  \left[  \frac{N}{2}\right]  \right)  }\left(  \frac{b+1}%
{2};\underset{\left[  \frac{N}{2}\right]  }{\underbrace{-1,...,-1}}%
;a+\frac{b+3}{2};\frac{1}{x_{1}^{2}},...,\frac{1}{x_{\left[  \frac{N}%
{2}\right]  }^{2}}\right)  \label{new}%
\end{equation}
where $x_{k}$ is defined in Eq.(\ref{root}). We believe that the identity in
Eq.(\ref{new}) derived from string theory was not known previously in the
literature \cite{jack}. In conclusion, we have shown that each amplitude of
process (A) can be expressed in terms of a single Lauricella function with
nonpositive integer $\beta_{j}$ and thus is a LSSA. As a result, all
scattering of string at arbitrary mass levels from D-brane calculated in (A)
form a part of an infinite dimensional representation of the exact $SL(K+3,%
\mathbb{C}
)$ symmetry of the bosonic string theory.

\section{Closed string decays into two open string}

In this section, we consider process (B), namely, closed string decays into
two open strings on the brane. We will adapt the same strategy used in the
last section and calculate only a typical term of a given process. We begin
with the kinematics of the decay process. The momentum conservation on the
D-brane reads%
\begin{equation}
\frac{1}{2}(k_{c}+D\cdot k_{c})+k_{1}+k_{2}=0 \label{11}%
\end{equation}
where $k_{c}$ is the momentum of the closed string state. In the usual
three-point amplitudes, momentum conservation completely constrains the
kinematics. In the presence of D-brane, the non-conservation of momentum in
the directions transverse to the D-brane gives precisely one kinematic
variable which can be defined to be%
\begin{equation}
t=-\left(  k_{1}+k_{2}\right)  ^{2}. \label{12}%
\end{equation}
By using Eq.(\ref{11}) and Eq.(\ref{12}), one easily gets%
\begin{align}
k_{1}\cdot k_{c}  &  =k_{1}\cdot D\cdot k_{c}=\frac{t+M_{1}^{2}-M_{2}^{2}}%
{2},\nonumber\\
k_{2}\cdot k_{c}  &  =k_{2}\cdot D\cdot k_{c}=\frac{t+M_{2}^{2}-M_{1}^{2}}{2},
\end{align}
which give
\begin{equation}
t=k_{1}\cdot k_{c}+k_{2}\cdot D\cdot k_{c}=k_{2}\cdot k_{c}+k_{1}\cdot D\cdot
k_{c}. \label{tt}%
\end{equation}
\qquad\qquad

We first calculate the amplitude of a closed string tachyon to decay into two
open string tachyons
\begin{align}
A_{tach}  &  =\int dx_{1}dx_{2}d^{2}z\left\langle e^{ik_{1}\cdot X\left(
x_{1}\right)  }e^{ik_{2}\cdot X\left(  x_{2}\right)  }e^{ik_{c}\cdot X\left(
z\right)  }e^{ik_{c}\cdot\bar{X}\left(  \bar{z}\right)  }\right\rangle
\nonumber\\
&  =\int dx_{1}dx_{2}d^{2}z\cdot\left(  x_{1}-x_{2}\right)  ^{k_{1}\cdot
k_{2}}\left(  z-\bar{z}\right)  ^{k_{c}\cdot D\cdot k_{c}}\left(
x_{1}-z\right)  ^{k_{1}\cdot k_{c}}\nonumber\\
&  \cdot\left(  x_{1}-\bar{z}\right)  ^{k_{1}\cdot D\cdot k_{c}}\left(
x_{2}-z\right)  ^{k_{2}\cdot k_{c}}\left(  x_{2}-\bar{z}\right)  ^{k_{2}\cdot
D\cdot k_{c}}.
\end{align}
The next step is to use $\{z_{1}$, $z_{2}$, $z_{3}$, $\bar{z}_{3}\}=\{-x$,
$x$, $i$, $-i\}$ to fix the $SL\left(  2,R\right)  $ invariance and obtain
\begin{align}
A_{tach}  &  =\int dx\text{ }x^{k_{1}\cdot k_{2}}\left(  x+i\right)
^{k_{1}\cdot k_{c}+k_{2}\cdot D\cdot k_{c}}\left(  x-i\right)  ^{k_{2}\cdot
k_{c}+k_{1}\cdot D\cdot k_{c}}\nonumber\\
&  =\int dx\text{ }x^{k_{1}\cdot k_{2}}\left(  x+i\right)  ^{t}\left(
x-i\right)  ^{t}\nonumber\\
&  =\int dx\text{ }x^{b_{0}}\left(  1+x^{2}\right)  ^{a_{0}}%
\end{align}
where we have used Eq.(\ref{tt}) and defined
\begin{equation}
a_{0}=t,b_{0}=k_{1}\cdot k_{2}. \label{tach}%
\end{equation}

We now turn to the general mass level case. We will again use the
\textit{solvability} of the LSSA discussed above to simplify the calculation
as before. A typical term of an arbitrary massive closed string state decays
into two arbitrary massive open string states can be written as%
\begin{equation}
A=\int dx\text{ }x^{k_{1}\cdot k_{2}+n_{b}}\left(  x+i\right)  ^{t+n_{a}%
}\left(  x-i\right)  ^{t+n_{c}}%
\end{equation}
where $n_{a}$, $n_{b}$ and $n_{c}$ are related to mass levels of $k_{c}$,
$k_{1}$ and $k_{2}$. At this point, we expect after summing up all terms in
the calculation, a real amplitude will be obtained. So we are going to
calculate only the real part of $A$%
\begin{align}
&  A+\bar{A}=\int_{-\infty}^{+\infty}dx\text{ }x^{k_{1}\cdot k_{2}+n_{b}%
}\left(  x^{2}+1\right)  ^{t}\left[  \left(  x+i\right)  ^{n_{a}}\left(
x-i\right)  ^{n_{c}}+\left(  x-i\right)  ^{n_{a}}\left(  x+i\right)  ^{n_{c}%
}\right] \nonumber\\
&  =\int_{-\infty}^{+\infty}dx\text{ }x^{k_{1}\cdot k_{2}+n_{b}+N}\left(
x^{2}+1\right)  ^{t+\min\left\{  n_{a},n_{c}\right\}  }\left[  \left(
1-\frac{i}{x}\right)  ^{N}+\left(  1+\frac{i}{x}\right)  ^{N}\right]
\label{111}%
\end{align}
where $N\equiv\left\vert n_{c}-n_{a}\right\vert $, and see whether the final
answer is a Lauricella function. Eq.(\ref{111}) can be further reduced to%
\begin{align}
A+\bar{A}  &  =\int_{-\infty}^{+\infty}dx\text{ }x^{b}\left(  x^{2}+1\right)
^{a}\underset{m=0}{\overset{N}{\sum}}\binom{N}{m}\left[  1+\left(  -1\right)
^{m}\right]  \left(  \frac{i}{x}\right)  ^{m}\nonumber\\
&  =2\underset{n=0}{\overset{\left[  \frac{N}{2}\right]  }{\sum}}\binom{N}%
{2n}\left(  -1\right)  ^{n}\int_{-\infty}^{+\infty}dx\text{ }x^{b-2n}\left(
x^{2}+1\right)  ^{a} \label{112}%
\end{align}
where we have defined%
\begin{align}
a  &  =t+\min\left\{  n_{a},n_{c}\right\}  ,\nonumber\\
b  &  =k_{1}\cdot k_{2}+n_{b}+N,
\end{align}
which are higher mass level generalization of Eq.(\ref{tach}). We can use the
change of variable $y=\frac{x^{2}}{x^{2}+1}$ to perform the integral in
Eq.(\ref{112})%
\begin{align}
&  \int_{-\infty}^{+\infty}dx\text{ }x^{b}\left(  1+x^{2}\right)
^{a}\nonumber\\
&  =\frac{1}{2}\left[  1+\left(  -1\right)  ^{b}\right]  \int_{0}^{1}dy\text{
}y^{\frac{b}{2}-\frac{1}{2}}\left(  1-y\right)  ^{-a-\frac{b}{2}-\frac{3}{2}%
}\nonumber\\
&  =\frac{1}{2}\left[  1+\left(  -1\right)  ^{b}\right]  \frac{\Gamma\left(
\frac{b}{2}+\frac{1}{2}\right)  \Gamma\left(  -a-\frac{b}{2}-\frac{1}%
{2}\right)  }{\Gamma\left(  -a\right)  }.
\end{align}
Finally, we obtain%
\begin{equation}
A+\bar{A}=2\underset{n=0}{\overset{\left[  \frac{N}{2}\right]  }{\sum}}%
\binom{N}{2n}\left(  -1\right)  ^{n}\frac{1}{2}\left[  1+\left(  -1\right)
^{b}\right]  \frac{\Gamma\left(  \frac{b}{2}+\frac{1}{2}+n\right)
\Gamma\left(  -a-\frac{b}{2}-\frac{1}{2}-n\right)  }{\Gamma\left(  -a\right)
}. \label{113}%
\end{equation}
To derive a Lauricella function in Eq.(\ref{113}), we note that%
\begin{equation}
\frac{\Gamma\left(  \frac{b}{2}+\frac{1}{2}+n\right)  \Gamma\left(
-a-\frac{b}{2}-\frac{1}{2}-n\right)  }{\Gamma\left(  -a\right)  }%
=\frac{\left(  \frac{b}{2}+\frac{1}{2}\right)  _{n}\Gamma\left(  \frac{b}%
{2}+\frac{1}{2}\right)  \Gamma\left(  -a-\frac{b}{2}-\frac{1}{2}\right)
}{\left(  -1\right)  ^{n}\left(  a+\frac{b}{2}+\frac{3}{2}\right)  _{n}%
\Gamma\left(  -a\right)  }.
\end{equation}
So Eq.(\ref{113}) can be written as%
\begin{align}
A+\bar{A}  &  =2\left[  1+\left(  -1\right)  ^{b}\right]  \frac{\Gamma\left(
\frac{b}{2}+\frac{1}{2}\right)  \Gamma\left(  -a-\frac{b}{2}-\frac{1}%
{2}\right)  }{\Gamma\left(  -a\right)  }\underset{n=0}{\overset{\left[
\frac{N}{2}\right]  }{\sum}}\binom{N}{2n}\frac{\left(  \frac{b}{2}+\frac{1}%
{2}\right)  _{n}}{\left(  a+\frac{b}{2}+\frac{3}{2}\right)  _{n}}\nonumber\\
&  =2\left[  1+\left(  -1\right)  ^{b}\right]  B\left(  -a-\frac{b+1}{2}%
,\frac{b+1}{2}\right) \nonumber\\
&  \times_{3}F_{2}\left(  \frac{b+1}{2},-\left[  \frac{N}{2}\right]  ,\frac
{1}{2}-\left[  \frac{N}{2}\right]  ;a+\frac{b+3}{2};\frac{1}{2};1\right)
\nonumber\\
&  =2\left[  1+\left(  -1\right)  ^{b}\right]  B\left(  -a-\frac{b+1}{2}%
,\frac{b+1}{2}\right) \nonumber\\
&  \times F_{D}^{\left(  \left[  \frac{N}{2}\right]  \right)  }\left(
\frac{b+1}{2};\underset{\left[  \frac{N}{2}\right]  }{\underbrace{-1,...,-1}%
};a+\frac{b+3}{2};\frac{1}{x_{1}^{2}},...,\frac{1}{x_{\left[  \frac{N}%
{2}\right]  }^{2}}\right)  \label{decay}%
\end{align}
where we have used the identities in Eq.(\ref{ac}) and Eq.(\ref{new}).
Finally, we can use the solvability property of $F_{D}^{(\left[  \frac{N}%
{2}\right]  )}$ with nonpositive $\beta_{j}$ to argue that the final amplitude
after summing up all typical terms of the decay process is a LSSA.

Incidentally, we note that the factor in the first line of Eq.(\ref{decay})
\begin{equation}
\frac{1}{2}\left[  1+\left(  -1\right)  ^{b}\right]  \frac{\Gamma\left(
\frac{b}{2}+\frac{1}{2}\right)  \Gamma\left(  -a-\frac{b}{2}-\frac{1}%
{2}\right)  }{\Gamma\left(  -a\right)  }=\int_{-\infty}^{+\infty}dx\text{
}x^{b}\left(  1+x^{2}\right)  ^{a}%
\end{equation}
with $a=2t-1,b=-2t$ can be calculated to be%

\begin{equation}
\frac{1}{2}\left[  1+\left(  1\right)  ^{-t}\right]  \frac{\Gamma\left(
-t+\frac{1}{2}\right)  \Gamma\left(  -2t+1+t-\frac{1}{2}\right)  }%
{\Gamma\left(  -2t+1\right)  }=\frac{\Gamma^{2}\left(  -t+\frac{1}{2}\right)
}{\Gamma\left(  -2t+1\right)  }. \label{22}%
\end{equation}
By using the duplication formula for the gamma function%
\begin{equation}
\Gamma\left(  -t+\frac{1}{2}\right)  =\frac{2^{1+2t}\sqrt{\pi}\Gamma\left(
-2t\right)  }{\Gamma\left(  -t\right)  },
\end{equation}
the result in Eq.(\ref{22}) can be further reduced to%
\begin{equation}
\frac{\Gamma^{2}\left(  -t+\frac{1}{2}\right)  }{\Gamma\left(  -2t+1\right)
}=\frac{2^{2+4t}\pi\Gamma\left(  -2t\right)  }{-2t\Gamma^{2}\left(  -t\right)
}=\frac{4\cdot16^{t}t\pi\Gamma\left(  -2t\right)  }{-2\left(  -t\Gamma\left(
-t\right)  \right)  ^{2}}=-2t\cdot16^{t}\frac{\pi\Gamma\left(  -2t\right)
}{\Gamma^{2}\left(  -t+1\right)  }.
\end{equation}
The factor $\frac{\Gamma\left(  -2t\right)  }{\Gamma^{2}\left(  -t+1\right)
}$ can also be found in \cite{Klebanov,D3} for the massless string/D-brane
decay process.

\section{Conclusion}

In conclusion, in this paper we have shown that each amplitude of processes
(A) and (B) for arbitrary massive string/D-brane states can be expressed in
terms of a single Lauricella function with nonpositive integer $\beta_{j}$ and
thus is a LSSA. To obtain the final results, we have used the
\textit{solvability} of the LSSA with nonpositive integer $\beta_{j}$ to
simplify the calculation.

In addition to the scattering processes calculated in (A), all decay
amplitudes of string/D-brane states at arbitrary mass levels calculated in (B)
also form a part of an infinite dimensional representation of the exact
$SL(K+3,%
\mathbb{C}
)$ symmetry of the bosonic string theory. The results in this paper extends
the previous exact $SL(K+3,%
\mathbb{C}
)$ symmetry of tree-level open bosonic string theory to include the D-brane.

\begin{acknowledgments}
This work is supported in part by the Ministry of Science and Technology
(MoST) and S.T. Yau center of National Yang Ming Chiao Tung University (NYCU), Taiwan.
\end{acknowledgments}

\end{document}